\documentclass[twoside,12pt]{article}
\usepackage{epsfig}

\def\Journal#1#2#3#4{{#1} {#2} (#4) #3 }

\def\NPA{{\em Nucl. Phys.} A}
\def\NPB{{\em Nucl. Phys.} B}

\def\NPB{{\em Nucl. Phys.} B}

\def\PLB{{\em Phys. Lett.} B}

\def\PRL{\em Phys. Rev. Lett.}
\def\JETP{\em Sov. Phys. JETP}

\def\PRD{{\em Phys. Rev.} D}
\def\PRC{{\em Phys. Rev.} C}

\def\Prog{\em Prog. Theor. Phys.}
\def\JHEP{\em JHEP}
\def\PhysRep{\em Rept. Prog. Phys}

\def\JPG{{\em  Journ. Phys. G}}
\def\NIM{\em Nucl.\ Instrum.\ Meth.\  A}

\newcommand{\be}{\begin{equation}}
\newcommand{\ee}{\end{equation}}
\newcommand{\bea}{\begin{eqnarray}}
\newcommand{\eea}{\end{eqnarray}}

\begin{document}

\title{ \vspace{1cm} Search for CP violation in the lepton sector}
\author{C. Volpe,$^{1}$
\\
$^1$Institut de Physique Nucl\'eaire, F-91406 Orsay cedex, \\
CNRS/IN2P3 and University of Paris-XI, France}
\maketitle
\begin{abstract} 
One of the major open issues in neutrino physics is the possible existence of CP violation in the neutrino sector.
Such an observation would have an important impact in various domains of physics, from high energy physics to cosmology. Its search requires future accelerator neutrino facilities producing intense and pure neutrino beams such as "beta-beams". Here we review the different beta-beam scenarios proposed so far and discuss the present status, with a particular emphasis on the original baseline scenario and its feasibility. Alternative strategies for the CP violation search are to be pursued as well. A possibility is to search for CP violation effects in astrophysical environments. Here we present recent analytical and numerical results obtained in the context of core-collapse supernovae. In particular, we point out the conditions under which there can be CP violating effects in dense media and show numerical results on the supernova (anti-)neutrino fluxes and on the electron fraction, relevant for the r-process nucleosynthesis. 
\end{abstract}

\section{Introduction}
In the last decade a series of experiments have provided us with essential information on neutrino properties, giving clear evidence for physics beyond the Standard Model. Neutrino oscillations first proposed by B. Pontecorvo \cite{Pontecorvo57}, are a firmly established phenomenon and most of the oscillation parameters have now been measured with accuracy. The ensemble of the present experimental results can be
interpreted in the context of three active flavours. In particular, the characteristic $L/E_{\nu}$ dependence\footnote{L is the distance between the neutrino source and the detector, $E_{\nu}$ is the neutrino energy.} has been identified as well both by the Super-Kamiokande and the Kamland experiments. The oscillation phenomenon implies that the flavour and the mass basis are related by a unitary matrix, the Maki-Nakagawa-Sakata-Pontecorvo matrix \cite{maki} :
\be\label{e:MNSP}
\label{e:U}
U  = \left(\matrix{
     1 & 0 & 0  \cr
     0 &  c_{23}  & s_{23} \cr
     0 & - s_{23} &  c_{23} }\right)
 \left(\matrix{
     c_{13} & 0 &  s_{13} e^{-i\delta}\cr
     0 &  1 & 0 \cr
     - s_{13} e^{i\delta} & 0&  c_{13} }\right)
 \left(\matrix{
     c_{12} & s_{12} &0 \cr
     - s_{12} & c_{12} & 0 \cr
     0 & 0&  1 }\right)
     \left(\matrix{
     1 & 0 &0 \cr
     0 & e^{i\alpha} & 0 \cr
     0 & 0&   e^{i\beta}}\right) ,
\ee 
with $c_{ij} = $cos$ \theta_{ij}$ ($s_{ij} = $sin$ \theta_{ij}$) and $\theta_{12}$, $\theta_{23}$ and $\theta_{13}$ the three neutrino mixing angles. Two angles and two squared mass differences have been precisely measured, namely $\delta m^2_{12}= 8 \times 10^{-5}$eV$^2$, $|\delta m^2_{23}|= 3 \times 10^{-3}$eV$^2$, sin$^{2} 2\theta_{12}=0.83$ and sin$^{2} 2\theta_{23}=1$ 
thanks to the solar and reactor experiments on one hand and to the athmospheric ones on the other
\cite{Amsler:2008zzb}.

Four important neutrino properties remain unknown : the value of the third mixing angle,
the absolute neutrino mass scale and hierarchy, the (Dirac versus Majorana) neutrino nature and the possible existence
of CP violation in the lepton sector. Upcoming experiments and projects that are at present at the R \& D level will address these questions. 
For the value of $\theta_{13}$ we only have an upper limit given by the Chooz experiment  \cite{Apollonio:1999ae}. In the near future three reactor experiments -- Double-Chooz \cite{Ardellier:2004ui}, RENO \cite{Kim:2008zzb}, Daya Bay \cite{Guo:2007ug} -- and the first super-beams experiments (T2K and NO$\nu$A) will be able to measure its value if  sin$^2 2 \theta_{13} < 0.02$ \cite{Huber:2009cw}. 
Note that the combination of the available experimental data gives indication that 
$ \theta_{13}$ might be close to the present Chooz limit \cite{Balantekin:2008zm,Fogli:2008jx}.
The second unknown parameter, relevant for oscillation experiments, is the Dirac phase $\delta$. 
Only if the third neutrino mixing angle is different from zero
there can be CP violating effects coming from the Dirac phase.
Indeed for a non-zero $\delta$ the MNSP matrix becomes complex, introducing a difference between neutrinos and anti-neutrinos. The matrix in Eq.(\ref{e:MNSP}) is completely specified if the Majorana phases $\alpha,\beta$ are also determined through the measurement of neutrinoless double-beta decay that addresses the crucial issue of the neutrino nature and of the (effective electron) neutrino mass. 
It is clear that the upcoming experiments will tell us if $\theta_{13}$ is
close to the present Chooz limit or if it is (very) small.
This will set our future strategy for CP searches with accelerator experiments.  

As is well known the reactor experiments are not sensitive to the Dirac phase since they measure electron (anti-)neutrino disappearance and the electron neutrino survival probability in vacuum does not depend on $\delta$. To this aim one needs appearance experiments
such as super-beam ones where oscillations of $\nu_{\mu}$ to $\nu_e$ are looked for.
As pointed out in  \cite{Huber:2009cw}, if sin$^2 2 \theta_{13} < 0.02$ the upgrades of the first generation super-beams (NO$\nu$A and T2K) have a CP discovery reach for a significant fraction of $\delta$ values at $90 \%$ CL while a $3\sigma$ discovery can be found only for a tiny fraction of the phase values. Therefore, unless we are in a very lucky situation in which the $\delta$ value is fixed at one of the few values that can be covered with these facilities, the measurement of the Dirac phase will require intense neutrino beams, based on new concepts, -- the neutrino factory or the "beta-beams" --, or the second generation super-beams. The neutrino factory exploits neutrino beams produced by the decay of stored muons and sent to far detectors. The beta-beam concept, first proposed by P. Zucchelli \cite{Zucchelli:2002sa}, uses the beta-decay of boosted radioactive ions. The goal of these facilities is to search for small $\theta_{13}$ values, determine the Dirac phase and identify the mass hierarchy. Here we discuss the different beta-beams scenarios proposed -- the original, the high energy scenarios
\cite{BurguetCastell:2003vv} as well as the electron capture variant \cite{Bernabeu:2005jh} -- and discuss some of the feasibility aspects. (For a review of all scenarios see  \cite{Volpe:2006in}.)
We will also mention low energy beta-beams \cite{Volpe:2003fi}, with neutrino energies in the 100 MeV energy range. The availability of such beams would broaden the physics case since they offer the unique opportunity to perform neutrino interaction studies of interest for nuclear structure, for the study of fundamental interactions and for the physics of core-collapse supernovae.

In paraller with the feasibility and physics reach studies of 
long term facilities such as neutrino factories and beta-beams,
it is important to explore complementary avenues e.g. by searching for indirect CP violation effects in astrophysical environments.
So far few attempts have been performed
in this direction. In \cite{Minakata:1999ze} the authors have studied the effects
of the Dirac phase in a star like our Sun. In particular, it has
been shown that the electron neutrino survival probability does not
depend on the CP phase even in matter, while the effects coming
from radiative corrections are expected to be extremely small.
In \cite{Akhmedov:2002zj} CP effects in core-collapse supernovae are considered in the extreme density limit. A somewhat different perspective is followed in \cite{Winter:2006ce} where
it is pointed out that the combination of an early CP measurement with Earth
based experiments and with neutrinos from astrophysical sources can significantly
improve the CP discovery reach. Recently we have been exploring 
the conditions under which there can be CP violating effects in supernovae \cite{Balantekin:2007es}. 
Exact analytical results have been obtained, for the first time, that are valid for any density profile. Besides, extensive numerical calculations have been performed quantifying possible
effects on the neutrino fluxes in the star, in an observatory on Earth and on the r-process nucleosynthesis. These results have been extended in a following work \cite{Gava:2008rp} including in particular the neutrino-neutrino interaction. The works in \cite{Balantekin:2007es,Gava:2008rp} have set the basis for the exploration of CP violating effects in dense media.

\begin{figure}[tb]
\begin{center}
\begin{minipage}[t]{8 cm}
\epsfig{file=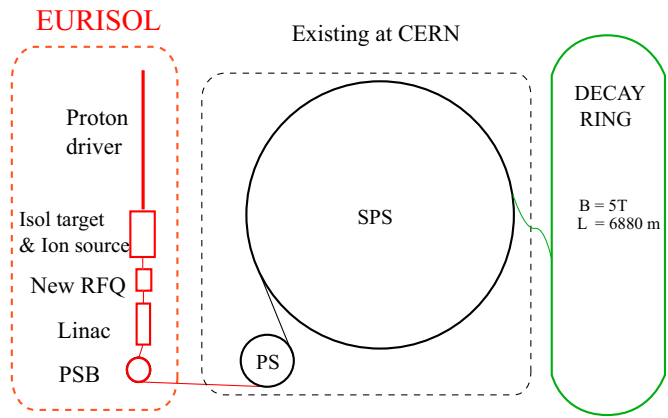,scale=0.4}
\end{minipage}
\begin{minipage}[t]{16.5 cm}
\caption{The beta-beam scenario based at CERN using existing accelerators (the PS and the SPS), in conjunction with the EURISOL facility, as first proposed by Zucchelli \cite{Zucchelli:2002sa}. The low energy accelerator infrastructures are according the feasibility study performed within the EURISOL (European Isotope Separator On-Line Radioactive Ion Beam Facility) Design Study (6th Framework Programme) \cite{eurisol}. The neutrino beams produced in the storage ring straight sections are sent to a large-size detector, such as MEMPHYS, located 130 km from CERN\label{fig1}.}
\end{minipage}
\end{center}
\end{figure}

\section{Beta-beams}

The CP violation search will require neutrino beams with the highest possible intensities, tiny intrinsic backgrounds and a very good control of systematic errors.
The beta-beam concept has three important advantages. The neutrino beams are pure in flavour since only electron neutrinos or anti-neutrinos can be produced, depending on the ion that decays through $\beta^+$ or $\beta^-$. This means that there is no beam related background. The neutrino intensity and energy spectrum is precisely known, since the number of ions is perfectly controlled.  The flux emittance is inversely proportional to the Lorentz boost factor $\gamma$.

\begin{figure}[tb]
\begin{center}
\begin{minipage}[t]{8 cm}
\epsfig{file=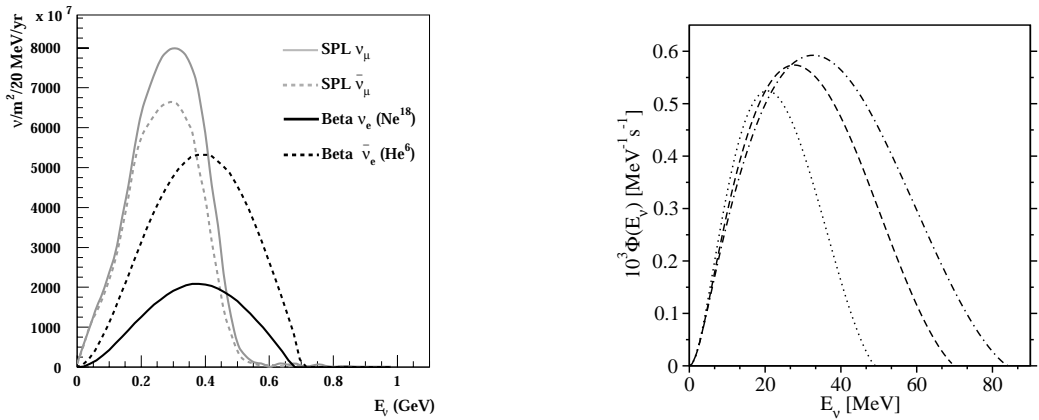,scale=0.52}
\end{minipage}
\begin{minipage}[t]{8 cm}
\epsfig{file=fluxes.eps,scale=0.3}
\end{minipage}
\begin{minipage}[t]{16.5 cm}
\caption{Left : Comparison of neutrino fluxes from a super-beam (SPL) and the standard beta-beam
with $\gamma=100$ \cite{Guglielmi05}.
Right : Electron anti-neutrino fluxes from low energy beta-beams, produced via the decay of
$^{6}$He ions boosted to $\gamma=7$ (dotted line), $\gamma=10$
(broken line) and $\gamma=12$ (dash-dotted line) \cite{Balantekin:2005md}.  \label{fig:nuflux}}
\end{minipage}
\end{center}
\end{figure}

\begin{figure}[tb]
\begin{center}
\begin{minipage}[t]{8 cm}
\epsfig{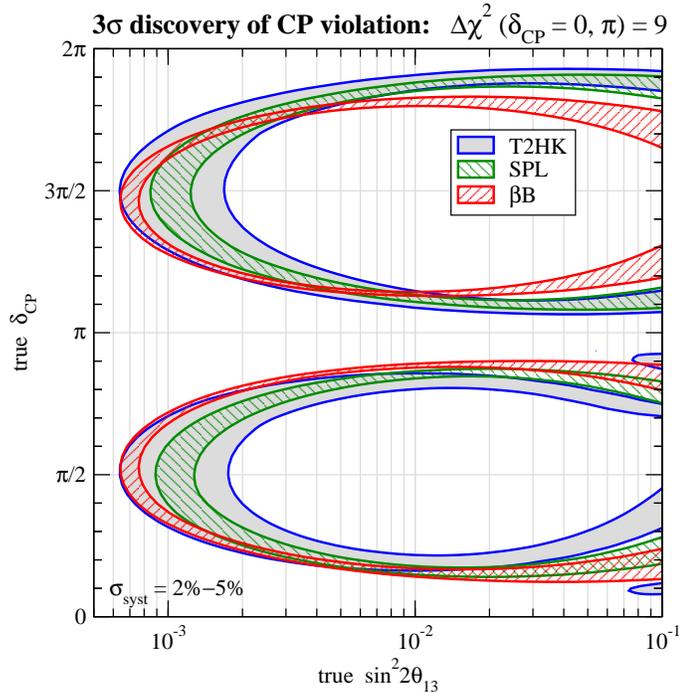}
\end{minipage}
\begin{minipage}[t]{16.5 cm}
\caption{Sensitivity on $\delta$ at 
  $3\sigma$ for
  the ($\gamma=100$) beta-beam ($\beta$B), a super-beam (SPL), 
  and T2HK (see text) as a function of the true value of sin$^2 2 \theta_{13}$. The width of the bands corresponds
  to values with 2$\%$ to 5$\%$ systematical errors. The
  dashed curves correspond to the combination of the beta-beam and the super-beam with
  10~yrs measurement time each and 2$\%$ systematical error. For parameter values inside the ellipse-shaped curves CP conserving
   values of $\delta_{\rm{CP}}$ can be excluded at $3\sigma$ $(\Delta\chi^2>9)$  
\cite{Campagne:2006yx}.\label{fig:CPs}}
\end{minipage}
\end{center}
\end{figure}

\begin{figure}[tb]
\begin{center}
\begin{minipage}[t]{8 cm}
\epsfig{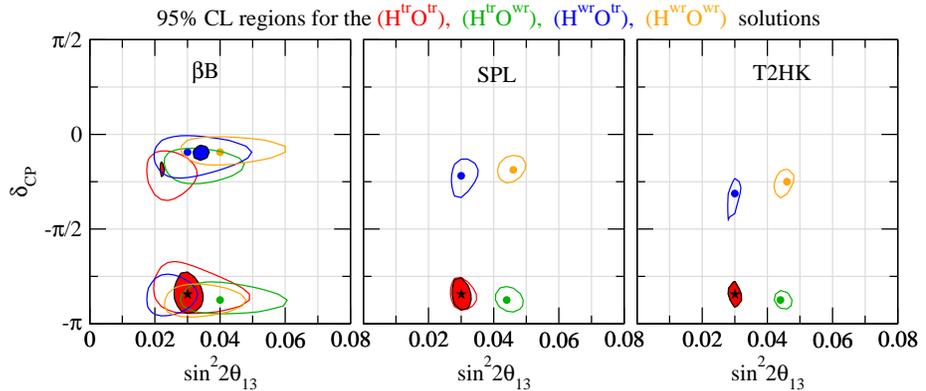}
\end{minipage}
\begin{minipage}[t]{16.5 cm}
\caption{The mass hierarchy and the octant degeneracy can be identified in the standard beta-beam scenario by exploiting the synergy with athmospheric neutrinos. 
Three cases are shown : the $\gamma=100$ beta-beam, the super-beam (SPL) at CERN and T2K phase II to Hyper-K. 
  The figure shows the
  allowed regions in sin$^22\theta_{13}$ and
  $\delta_\mathrm{CP}$ for accelerator data alone (contour lines) and accelerator plus atmospheric
  data combined (colored regions). Solution for the true/wrong mass hierarchy (octant of
  $\theta_{23}$) are indicated with
  $\mathrm{H^{tr/wr} (O^{tr/wr})}$.
  The true parameter values are 
$\delta_\mathrm{CP} = -0.85 \pi$, sin$^22\theta_{13} = 0.03$, sin$^2\theta_{23} =0.6$
\cite{Campagne:2006yx}.\label{fig:deg}}
\end{minipage}
\end{center}
\end{figure}

The beta-beam scenarios proposed so far can be summarized as follows \cite{Volpe:2006in} :
\begin{itemize}
\item {\bf the standard beta-beam scenario \cite{Zucchelli:2002sa} :}
In the original baseline scenario \cite{Zucchelli:2002sa}, the beta-beam facility is hosted at CERN (Figure \ref{fig1}). The ions are produced through the ISOL technique, bunched and accelerated first at several hundred MeV per nucleon, then injected in the PS and SPS to reach $\gamma=100$ (Figure \ref{fig:nuflux}). Finally the ions are stored in a storage ring with long straight sections (2.5 km for a total length of 7 km) that point to a (large size) water \v{C}erenkov detector, located in an (upgraded) Underground Fr\'ejus laboratory. The search for CP effects is performed through the comparison of $\nu_e \rightarrow \nu_{\mu}$ versus $\bar{\nu}_e \rightarrow \bar{\nu}_{\mu}$ oscillations.  Note that if one sends a super-beam to the same detector, T breaking can also be studied by measuring  $\nu_{\mu} \rightarrow \nu_e $ oscillations as well as CPT breaking through the comparison of $\nu_e \rightarrow \nu_{\mu}$ and $\bar{\nu}_{\mu} \rightarrow \bar{\nu}_e$ oscillations. The discovery reach for CP violation of the original scenario has been first investigated in \cite{Mezzetto:2003mm} and in detail in \cite{Campagne:2006yx}.
With such a baseline one has a sensitivity on the third neutrino mixing angle of
sin$^2 \theta_{13} \geq 5 \cdot 10^{-4}$ (3$\sigma$ for 75 $\%$ of the values of $\delta$) and has a CP violation discovery reach at 3$\sigma$ for sin$^2 \theta_{13}$ values down to 
$7 \cdot 10^{-4}$ (Figure \ref{fig:CPs}).
While the baseline is too short to exploit the matter effects in order to determine the mass hierarchy, this goal can be achieved by using a combination of the neutrino beams with the athmospheric ones as first proposed in \cite{Huber:2005ep} and used in \cite{Campagne:2006yx} to show that the octant and the mass hierarchy identification can be determined with the standard beta-beam (Figure \ref{fig:deg}).

\item {\bf the low-energy beta-beam \cite{Volpe:2006in} : } This corresponds to ions running at $\gamma = 5-15$ to produce neutrinos in the 100 MeV energy range (Figure \ref{fig:nuflux}). 
Such beams would allow : i) neutrino scattering measurements on nuclei, to improve our knowledge of the isospin and spin-isospin nuclear response \cite{Volpe:2006in,Serreau:2004kx,McLaughlin:2004va,Lazauskas:2007bs} for various timely applications (for example to costrain neutrinoless double-beta decay calculations \cite{Volpe:2005iy}, 
r-process calculations and the detection of (relic) supernova neutrinos; ii) searches of physics beyond the Standard Model (through e.g. the measurement of the Weinberg angle \cite{Balantekin:2005md}, a new Conserved Vector Current hypothesis test \cite{Balantekin:2006ga},
of coherent neutrino-nucleus scattering measurement \cite{Bueno:2006yq}, searches of non-standard contributions \cite{Barranco:2007tz}); iii) core-collapse supernova physics studies \cite{Volpe:2006in,Jachowicz:2006xx,Jachowicz:2008kx}. This option might require a devoted storage ring \cite{Serreau:2004kx} or could use the low energy neutrino beam component of the standard beta-beam by putting one/two detectors at off-axis \cite{Lazauskas:2007va}. 

\item {\bf the high energy beta-beam  \cite{BurguetCastell:2003vv} :} In the first proposal 
the ions run at higher $\gamma$ (about 350) and the baseline is increased accordingly to about 700 km. Such a long baseline is sensitive to the neutrino hierarchy through matter effects. Note that increasing the boost factor require a bigger storage ring since the length increases linearly with $\gamma$. On the other hand a higher $\gamma$ produces higher rates in the detector for the same ion intensity. The work in \cite{BurguetCastell:2003vv} has stimulated numerous studies (see e.g. \cite{BurguetCastell:2005pa,Huber:2005jk,Donini:2006dx,Coloma:2007nn,Choubey:2009ks,Agarwalla:2008ti}).
For example in \cite{Agarwalla:2007ai} the magic baseline from CERN to the India-based Neutrino Observatory (7152 km) is investigated. At such a distance the $\nu_e$ to $\nu_{\mu}$ survival probability 
has no dependence on $\delta$ and allows to measure the neutrino hierarchy without any degenerate solution. A beta-beam experiment from FNAL to DUSEL is proposed in \cite{Agarwalla:2009ki}.
Boosting the $\nu_e$ and $\bar{\nu}_e$ to the maximum energy possible with the Tevatron
one could exploit an interesting combination of the first and the second maximum that helps
reducing the parameter degeneracies.  
 
\item {\bf the electron capture option  \cite{Bernabeu:2005jh} :} Here the ions decay by electron capture instead of beta-decay, producing electron neutrinos only. The advantage of such beams is that the neutrinos are monoenergetic. On the other hand this option requires acceleration and storage of partially stripped ions, which is quite challenging from the technical point of view. Candidate ions have rather long lifetimes like for example $^{150}$Dy. Promising sensitivities on CP can be obtained by running the ions at two different $\gamma$ tuned to the first and second oscillation maximum \cite{Bernabeu:2007sx}. A combination of a beta- and electron-capture beam is investigated in \cite{Bernabeu:2009np} considering different baselines.

\end{itemize}

\begin{figure}[tb]
\begin{center}
\begin{minipage}[t]{8 cm}
\epsfig{file=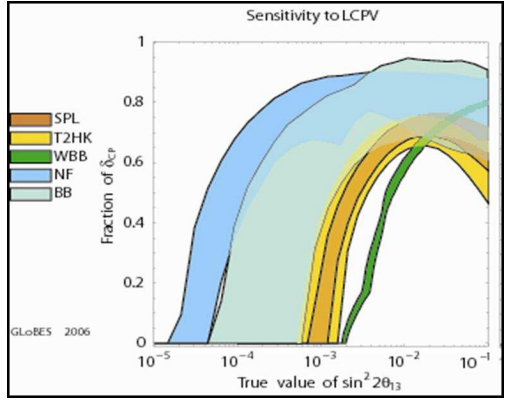,scale=0.5}
\end{minipage}
\begin{minipage}[t]{16.5 cm}
\caption{The CP discovery reach for various proposed long-term accelerator facilities : super-beams 
(T2HK in yellow, SPL in orange, WBB in green), beta-beams (light green band) and a neutrino factory (blue band). The sensitivities are shown for the fraction of all possible values of the true value of the CP phase $\delta$. Each band is determined by two different options for each facility,
a conservative one giving the lower end while an optimised one gives the upper end (see text)
\cite{Bandyopadhyay:2007kx}.\label{fig:isscp}}
\end{minipage}
\end{center}
\end{figure}

The physics reach of the various beta-beam scenarios has been investigated in great detail in the last years. The studies performed can be classified in two categories, based on what one can call the "conservative" or "optimized" attitude. In the former the ion intensities and boosts are obtained extrapolating well known technologies and existing accelerator infrastructures (the present PS and SPS at CERN for example). In the latter the ion intensities and boosts are treated as "free" parameters, with the aim of exploring the conditions to achieve optimal sensitivities.
Figure \ref{fig:isscp} presents a comparison on the CP discovery reach of the future long-term facilities \cite{Bandyopadhyay:2007kx}. For each option the figure shows the sensitivity corresponding both to the conservative and to the optimized options. The T2HK (SPL) cases correspond to a 4 MW proton driver, with the neutrino beams firing to a megaton class (440 kt) water \v{C}erenkov detector located at 295 (130) km distance. The running time is 2 (8) years for neutrinos (-neutrinos). The conservative option for the beta-beam is the CERN-Fr\'ejus scenario with $\gamma=100$, while
the optimized one is with a $\gamma=350$ and a baseline of 700 km. In both cases a 440 kt water \v{C}erenkov detector is considered and the running time is 5 (5) years with 2.9 10$^{18}$ (1.1 $10^{18}$) $^{6}$He ($^{18}$Ne) per year. For the neutrino factory, the conservative (optimised) setup uses
10$^{21}$ muon decays per year, with a stored muon-beam energy of 50 (20) GeV, The beams point to a (two) 50 kton detectors located at 4000 (4000 and 7500) km. The running time is 4 (5) years with
$\mu^+$ and $\mu^-$. As it can be seen from Figure \ref{fig:isscp} if sin$^2 2\theta_{13}>$0.02 the three options have very similar sensitivities on $\delta$. For values of
5 10$^{-4} <$  sin$^2 2 \theta_{13} <$ 10$^{-2}$ the super-beams are outperformed by the beta-beam and the neutrino factory. Only the optimised neutrino factory can reach values smaller than sin$^2 2 \theta_{13} <$ 5 10$^{-4}$ \cite{Bandyopadhyay:2007kx}, while the optimised beta-beam and the conservative neutrino factory options have a comparable performance.

A first feasibility study of the original beta-beam scenario is performed in \cite{Autin:2002ms}, while a
detailed investigation has just been completed within the EURISOL Design Study \cite{eurisol}. An important step forward has been made. Indeed most aspects appear under control (e.g. bunching of the ions, the stacking method in the storage and contamination issues). Reaching the required ion intensities will need further investigations \cite{Wildner:2008}. 
Two sets of ions have been discussed so far, namely $^{6}$He and $^{18}$Ne on one hand and $^{8}$Li and $^{8}$B on the other. The production of $^{6}$He and $^{18}$Ne through the standard ISOL technique as well as
the direct production method have been studied within the EURISOL DS.  
The production of $^{8}$Li and $^{8}$B with a storage ring method has been recently proposed \cite{Rubbia:2006pi} and will be investigated within the EUROnu initiative.  

\section{CP violation and astrophysics}
\subsection{The conditions for CP violation in core-collapse supernovae}
Our understanding of neutrino propagation in massive stars has been revolutionized in the last few years after the use of temporally evolving density profiles with shock waves and the inclusion of the neutrino-neutrino interaction.  This is very different from the solar case where the neutrino propagation is well understood in terms of the standard MSW paradigm. Indeed new phenomena have emerged, such as multiple resonances and phases effects on one hand, and of collective effects on the other. Besides, the study of the impact of turbulence is just at its beginning. (For a review see
\cite{Duan:2009cd}).

Recently we have investigated possible CP violating effects coming from the Dirac phase in supernovae.
Here we describe the conditions under which there can be CP effects. We use
the procedure established in \cite{Balantekin:2007es} and sketch the main steps of the demonstrations :
\begin{itemize}
\item within the standard MSW effect at tree level \cite{Balantekin:2007es}
\item with the neutrino-neutrino interaction term with/without
the $V_{\mu\tau}$ potential coming from the radiative contributions \cite{Gava:2008rp}.
\end{itemize}

\subsubsection{The neutrino evolution equations}
Let us 
consider the three flavour neutrino evolution equations in matter :
\begin{equation}
\label{msw}
i \frac{\partial}{\partial t} \left(\matrix{ \Psi_e\cr \Psi_{\mu} \cr
    \Psi_{\tau}} \right)  = \left[ T_{23}T_{13}T_{12} \left(\matrix{
     E_1 & 0 & 0  \cr
     0 &  E_2  & 0 \cr
     0 & 0 &  E_3 }\right)
T_{12}^{\dagger}T_{13}^{\dagger} T_{23}^{\dagger} +
\left(\matrix{
     V_c+V_n & 0 & 0  \cr
     0 &  V_n  & 0 \cr
     0 & 0 &  V_n + V_{\mu\tau} }\right)+ H_{\nu\nu} \right] \left(\matrix{ \Psi_e\cr \Psi_{\mu}
    \cr     \Psi_{\tau}} \right),
\end{equation}
where $V_c (x) = \sqrt{2} G_F  N_e (x)$ for the charged  and 
$V_n (x) =  - \frac{1}{\sqrt{2}} G_F N_n(x)$ 
for the neutral currents. Since $V_n$ only contributes an overall phase
to the neutrino evolution it can be ignored. The $V_{\mu\tau}$ term is due to radiative
corrections and is such that $V_{\mu\tau}/V_c \approx 10^{-5}$ \cite{Botella:1986wy}.  The last term of the Hamiltonian corresponds to the neutrino-neutrino contribution and is given by :
\begin{equation}
 \label{e:4}
H_{\nu \nu} = \sqrt{2} G_F
\sum_{\alpha} \sum_{\nu_{\alpha},\bar{\nu}_{\alpha}} \int
 \rho_{{\nu}_{\underline{\alpha}}} ({\bf q}')(1 - {\bf \hat{q}} \cdot {\bf \hat{q}'}) dn_{\alpha} dq' 
\end{equation}
where $G_F$ is the Fermi coupling constant, 
$\rho = \rho_{{\nu}_{\underline{\alpha}}}$  ($- \rho^*_{{\nu}_{\underline{\alpha}}}$) is 
the density matrix for neutrinos (anti-neutrinos), 
${\bf q}$ (${\bf q'}$) denotes the
momentum of the neutrino of interest (background neutrino) and $dn_{\alpha}$ is
the differential number density.  In the single-angle approximation, that assumes that the neutrinos
are all emitted with the same angle, i.e.  $ \rho ({\bf q}) = \rho (q)$, Eq.(\ref{e:4}) reduces to
\begin{equation}\label{e:4bis}
H_{\nu \nu} ={ \sqrt{2} G_F \over {2 \pi R_{\nu}^2}} D(r/ R_{\nu}) \sum_{\alpha} \int [\rho_{{\nu}_{\underline{\alpha}}} (q') L_{{\nu}_{\underline{\alpha}}} (q') - \rho_{\bar{{\nu}}_{\underline{\alpha}}}^*(q')  L_{\bar{{\nu}}_{\underline{\alpha}}}(q')] dq'
\end{equation}
with the geometrical factor $D(r/R_{\nu})$,
where the radius of the neutrino sphere is $R_{\nu}$, and
$L_{{\nu}_{\underline{\alpha}}}(q)$ are the neutrino fluxes at the neutrinosphere,
which can be taken Fermi-Dirac or power-law distributions.

\subsubsection{The conditions for CP violating effects}
\noindent
{\bf The standard MSW case}\\
To understand the origin of the CP effects, it is useful to rewrite Eq.(\ref{msw}) in the $T_{23}$ rotated basis :
\begin{eqnarray}
\label{rot1}
\tilde{\Psi}_{\mu} &=& \cos{\theta_{23}} \Psi_{\mu} -
\sin{\theta_{23}} \Psi_{\tau}, \\
\tilde{\Psi}_{\tau} &=& \sin{\theta_{23}} \Psi_{\mu} +
\cos{\theta_{23}} \Psi_{\tau},
\end{eqnarray}
If one neglects the neutrino-neutrino interacation, after some calculations one can show that the whole $\delta$ dependence of the Hamiltonian
can be factorized out as
\begin{equation}\label{e:7}
\tilde{U} ({\delta}) = S^{\dagger} \tilde{U}(\delta=0) S  \leftrightarrow \tilde{H}(\delta) = S^{\dagger} \tilde{H}(\delta=0) S
\end{equation}
where the CP dependence is in the unitary matrix
\begin{equation}
\label{s}
S^{\dagger} = \left(\matrix{
     1   & 0 & 0  \cr
     0 &  1    & 0 \cr
     0 & 0 &  e^{i\delta}   }\right) .
\end{equation}
From Eq.(\ref{e:7}) one can show that the following exact relations on the probabilities hold
\begin{equation}
\label{sq1}
P (\nu_e \rightarrow \nu_e, \delta \neq 0) =
P (\nu_e \rightarrow \nu_e, \delta =0)  .
\end{equation} 
and
\begin{equation}
\label{sq2}
P (\nu_{\mu} \rightarrow \nu_e, \delta \neq 0) + 
P (\nu_{\tau} \rightarrow \nu_e, \delta \neq 0) =
P (\nu_{\mu} \rightarrow \nu_e, \delta =0) + 
P (\nu_{\tau} \rightarrow \nu_e, \delta =0) .
\end{equation} 
The first relation was first shown in \cite{Minakata:1999ze} with a different procedure, in the context of solar neutrinos.  
Eq.(\ref{sq2}) is new and has important implications on the neutrino fluxes in a core-collapse supernova. We emphasize that our relations Eqs.(\ref{sq1}-\ref{sq2}) are exact and valid for any density profile. In a following work \cite{Kneller:2009vd}, these and new relations among the probabilities have been obtained using the adiabatic basis.

The consequence on the electron (anti-)neutrino fluxes can be immediately seen from
\begin{equation}\label{e:fluxnue}
{\phi}_{\nu_e}(\delta) =  L_{\nu_e}P(\nu_e \rightarrow \nu_e) + 
L_{\nu_{x}}(P(\nu_{\mu} \rightarrow \nu_e)+P(\nu_{\tau} \rightarrow \nu_e))
\end{equation}
where $\nu_{x}=\nu_{\mu}$ or $\nu_{\tau}$, which holds if and only if 
if the $\nu_{\mu}$ and $\nu_{\tau}$ neutrino fluxes are exactly equal at the neutrinosphere.
In this case ${\phi}_{\nu_e}(\delta)$ and ${\phi}_{\bar{\nu}_e}(\delta)$ do not depend on $\delta$.

\noindent
{\bf The case with the neutrino-neutrino interaction}\\
To show the origin of CP effects in this case, it is better to
start from the Liouville-Von Neumann equation for the density matrix ($\hbar=1$) :
\begin{equation}\label{e:13a}
i{{d \rho_{{\nu}_{\underline{\alpha}}}(\delta)} \over{dt}} = [U H_{vac} U^{\dagger} + H_m + H_{\nu \nu}(\delta),\rho_{{\nu}_{\underline{\alpha}}}(\delta)] ,
\end{equation}
with $H_{vac}$ and $H_m$ the usual vacuum and matter contributions.
To prove our result (that the CP-violating phase can be factorized out of the total Hamiltonian which includes $H_{\nu \nu}$), one has to rotate in the $T_{23}$ basis :
\begin{equation}\label{e:13b}
i{{d S\tilde{\rho}_{{\nu}_{\underline{\alpha}}}(\delta)S^{\dagger}} \over{dt}} = [T_{13}^0T_{12}H_{vac}T_{12}^{\dagger}{T_{13}^0}^{\dagger}+H_m+S\tilde{H}_{\nu \nu}(\delta)S^{\dagger},S\tilde{\rho}_{{\nu}_{\underline{\alpha}}}(\delta)S^{\dagger}] , 
\end{equation}\\
We now show that Eq.(\ref{e:7}) is indeed satisfied for the total Hamiltonian in Eq.(\ref{e:13a}) including the non-linear $H_{\nu \nu}$ term of Eq.(\ref{e:4}).
Let us now consider the evolution equation of the linear combination
$ \sum_{\nu_{\alpha}}L_{\nu_{\underline{\alpha}}} S\tilde{\rho}_{{\nu}_{\underline{\alpha}}}({\bf q},\delta)S^{\dagger} $ at a given momentum ${\bf q}$.
At the initial time, this quantity reads, in the $T_{23}$ basis as :
\begin{equation}\label{e:13c}
\sum_{\nu_{\alpha}}  L_{\nu_{\underline{\alpha}}}  S\tilde{\rho}_{{\nu}_{\underline{\alpha}}} ({\bf q},\delta, t=0)S^{\dagger} = \left(\matrix{
     L_{\nu_{\underline{e}}}  & 0 & 0  \cr
     0 &  c_{23}^2L_{\nu_{\underline{\mu}}}+s_{23}^2 L_{\nu_{\underline{\tau}}}  & c_{23}s_{23}e^{-i\delta}(L_{\nu_{\underline{\mu}}}-L_{\nu_{\underline{\tau}}}) \cr
     0 &  c_{23}s_{23}e^{i\delta}(L_{\nu_{\underline{\mu}}}-L_{\nu_{\underline{\tau}}}) &  s_{23}^2 L_{\nu_{\underline{\mu}}}+ c_{23}^2L_{\nu_{\underline{\tau}}}  }\right)
\end{equation} 
One immediately sees that this quantity does not depend on $\delta$ if and only if $ L_{\nu_{\underline{\mu}}}= L_{\nu_{\underline{\tau}}}$.
Moreover, one can show that since the total Hamiltonian of Eq.(\ref{e:13b}) is independent of $\delta$ at initial time (for $\tilde{H}_{\nu \nu}$ under the condition 
$L_{\nu_{\underline{\mu}}}= L_{\nu_{\underline{\tau}}}$),
this is true at any time, 
by recurrence from the Liouville-Von Neumann equation Eq.(\ref{e:13a}).
The same applies to the anti-neutrinos. 
One can see that, since $\tilde{H}_{\nu \nu}$ is proportional to the quantity 
in Eq.(\ref{e:13c}) (both for neutrinos and for anti-neutrinos) this implies that 
the factorization in Eq.(\ref{e:7}) holds
\begin{equation}\label{e:13d}
\tilde{H}_{\nu \nu}(\delta) = S \tilde{H}_{\nu \nu}(\delta=0) S^{\dagger}.
\end{equation}
Hence Eqs.(\ref{sq1}-(\ref{sq2}) are valid for the total Hamiltonian with the $\nu\nu$ term as well.
Note that this result is true independently on the angular assumption made for the neutrino-neutrino contribution (single-angle versus multi-angle).

\begin{figure}[tb]
\begin{center}
\begin{minipage}[t]{8 cm}
\epsfig{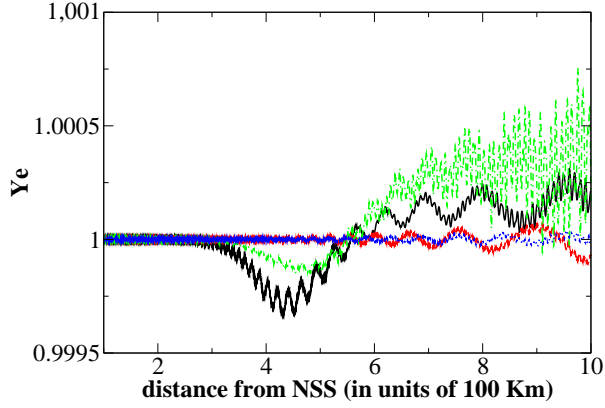}
\end{minipage}
\begin{minipage}[t]{16.5 cm}
\caption{Ratios of the electron fraction for $\delta=180^{\circ}$
compared to $\delta=0^{\circ}$,
as a function of the distance from the neutron-star surface. The
initial $\nu_{\mu},\nu_{\tau}$ fluxes have temperatures which differs
by 1 MeV (see text).  The results correspond to the normal hierarchy
and sin$^2 2\theta_{13}=0.19$ \cite{Balantekin:2007es}.\label{fig:ye}}
\end{minipage}
\end{center}
\end{figure}

\noindent
{\bf General conditions for CP violating effects}\\
From the above argument it is clear that there are at least  two possible sources of CP effects  in dense media : 1) any  contribution to the Hamiltonian that breaks the factorization Eq.(\ref{e:7}) in which case Eqs.(\ref{sq1}-\ref{sq2}) do not hold any more; 2) any physical effect that makes the $\mu$ and $\tau$ neutrino fluxes at the neutrinosphere differ, in which case the relation Eq.(\ref{sq2}) does not hold any more. 
It is easy to show that conditions 1)  is for example fulfilled when (standard or non-standard) radiative corrections are included. As a consequence the electron neutrino survival probability becomes dependent on $\delta$, breaking Eq.(\ref{sq1}). Such effect that is in principle very small, can be amplified in presence of the non-linear $\nu\nu$ contribution. Physics beyond the Standard Model, such as Flavour Changing Neutral Currents, can also break the condition 1) and 2) at the same time. This can induce indirect CP effects in supernovae.

\subsubsection{CP effects in core-collapse supernovae : Numerical results}
We have performed numerous calculations of the CP effects on the neutrino fluxes within a supernova \cite{Balantekin:2007es,Gava:2008rp}. The input parameters used are the oscillation best fit values and supernova density profiles that fit supernova simulations. We find significant CP effects on the muon and tau neutrino fluxes. However such effects do not have any impact since they disappear when one sums up the electron, muon and tau neutrino flux contributions since they all interact through neutral current. 

Concerning the possible CP effects on the nucleosynthesis of heavy elements (the r-process), Figure \ref{fig:ye}  shows the effects on the electron fraction that governs the neutron to proton ratio, a key parameter for the r-process abundance calculations. One sees that for this observable the effects appear to be
very small, although the effect on the electron (anti-)neutrino fluxes are of the order of several percent. The CP effect in an observatory on Earth comes out to be too small to be detectable. Such effects appear at high neutrino energies in electron anti-neutrino scattering on protons, the main detection channel of water \v{C}erenkov and scintillator detectors. Note that the size of the modifications due to the $\delta$ phase does not depend on the size of the observatory \cite{Balantekin:2007es}. These results have been obtained by including the coupling of neutrinos with matter only (standard MSW). 
\begin{figure}[tb]
\begin{center}
\begin{minipage}[t]{8 cm}
\epsfig{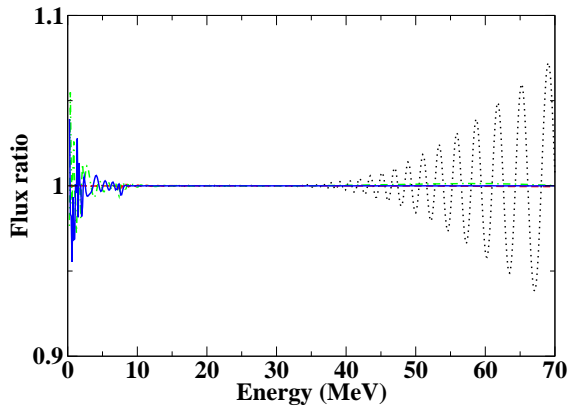}
\end{minipage}
\begin{minipage}[t]{16.5 cm}
\caption{Ratios of the $\nu_e$ fluxes for a CP violating phase $\delta=180^{\circ}$ over $\delta= 0^{\circ}$
as a function of neutrino energy, at 1000 km within the star, without the $\nu\nu$ interaction and radiative corrections (dotted), with the $\nu\nu$ interaction (dashed),
with the $\nu\nu$ interaction and radiative corrections taking different $\mu$ and $\tau$ fluxes (solid). 
The case where the $\nu\nu$ interaction and radiative corrections are taken into account with
$\nu_{\mu}$ and $\nu_{\tau}$ fluxes equal (dot-dashed) is also shown.
The results correspond to an inverted hierarchy and a small $\theta_{13}$ \cite{Gava:2008rp}.\label{fig:flux}}
\end{minipage}
\end{center}
\end{figure}
Further calculations should be performed to see how the CP effect on the electron
fraction as well as in an observatory on Earth can be modified when one includes the neutrino-neutrino interaction. 

Figure   \ref{fig:flux}
shows the CP effects on the neutrino fluxes in the star when the radiative corrections as well as the neutrino-neutrino interaction contribution are included as well \cite{Gava:2008rp}.
Several differences appear when this term is included. In particular, the synchronization
regime that is present in the inner part of the star tends to freeze the CP effects that only appear during the bipolar oscillations regime. Contrarily to the calculations without the $\nu\nu$ term,
it is the low energy neutrinos that are affected by the CP phase instead of the high energy ones. The neutrino fluxes in the ten MeV energy range are modified by approximately $5-10 \%$ when the $\nu_{\mu}$ and the  $\nu_{\tau}$ neutrino luminosities at the neutrinosphere
differ by a few percent (as an example). Such effects are also present if the mu and tau neutrino fluxes at the neutrinosphere are taken equal if radiative corrections are taken into account. Indeed the size of the CP effects is usually at the level of $\approx 10^{-4}$
$- 10^{-5}$ when the radiative corrections only are included. This is in agreement with what was found in \cite{Minakata:1999ze} in the context of solar neutrinos. It is the non-linearity of the neutrino evolution equations that amplifies such effect at the level of a few percent, as shown in Figure \ref{fig:flux} \cite{Gava:2008rp}. Note that non-standard contributions (such as those from SUSY models) could make the radiative corrections  much larger \cite{Gava:2009gt}. How much the CP effects can be amplified in this case requires further investigation.
 
\section{Conclusion}
Crucial open questions remain in neutrino physics, among which the value of the third neutrino mixing angle, the neutrino nature, the mass scale and the hierarrchy, and the possible existence  of CP violation in the lepton sector. If sin$^2 2 \theta_{13} > 0.02$
the third mixing angle will be measured by next reactor and the first super-beam experiments. However, unless nature has fixed the CP phase to one of the few values that can be covered with the upgrades of such super-beam facilities, the search for the CP violating Dirac phase will require long-term accelerator facilities, such as the second generation super-beams, the neutrino factory and the beta-beams. As far as beta-beams are concerned, many baseline scenarios have been proposed so far, for which the physics reach is under intense investigation. The CERN-Fr\'ejus baseline has now been studied in depth both from the point of view of the physics reach and of the feasibility, thanks to the EURISOL Design Study just ended (FP6). 
Further investigation of the ion intensity with the ISOL technique, the direct production and the storage ring methods need to be done/
The standard baseline is very attractive if $\theta_{13}$ is large or not too small. 
In particular, it can pin down if 
sin$^2 \theta_{13} \geq 5 \cdot 10^{-4}$ (3$\sigma$ for 75 $\%$ of the values of $\delta$) and has a CP violation discovery reach at 3$\sigma$ for sin$^2 \theta_{13}$ values down to 
$7 \cdot 10^{-4}$.
The octant degeneracy and the mass hierarchy can be identified as well by exploiting the synergy with athmospheric neutrinos in the same large size detector. 

It is important to pursue alternative searches for CP violation through the exploration of indirect effects e.g. in astrophysical contexts. 
Recently it has been demonstrated that there can be CP violating effects in core-collapse supernovae. These can emerge for example from radiative corrections, from physics beyond the Standard Model (e.g. from FCNC) or any contribution for which the CP dependence of the total Hamiltonian does not satisfy the factorization condition. Numerical results have shown effects up to 10 $\%$ on the neutrino fluxes in the star. The possible impact on observables like the electron-fraction, a key parameter for the r-process, requires further investigation.

\end{document}